\documentclass[letterpaper]{article}
\usepackage[centertags]{amsmath}
\usepackage{amssymb}
\usepackage[psamsfonts,mathscr]{eucal}
\usepackage[]{graphicx}
\usepackage[dvipsnames]{xcolor}
\usepackage{color}
\usepackage[normalsize,bf,tight]{subfigure}
\usepackage{enumitem, cases}
\usepackage[mathlines]{lineno}
\usepackage[]{upgreek}
\usepackage[utf8]{inputenc}
\usepackage{babel}
\usepackage[labelfont=bf,labelsep=space]{caption}
\usepackage[margin=1.0in]{geometry}
\usepackage[numbers,sort&compress]{natbib}
\definecolor{Blue}{rgb}{0,0,.8}
\definecolor{Green}{rgb}{0,.5,0}
\definecolor{Gray}{rgb}{.25,.25,.25}
\definecolor{Red}{rgb}{.8,0,0}
\usepackage[colorlinks=true,citecolor=Green,linkcolor=Red,urlcolor=Blue,bookmarks=false]{hyperref}

\usepackage[color=Thistle!70!White,textsize=scriptsize,textwidth=1.25in]{todonotes}

\graphicspath{{Paper_Figures/}{Figures_V2/}}

\title{\bf Damage Rate Laws and Failure Statistics for Lumped Coupled-Field Systems via Averaging}

\author{Arjun~Roy\thanks{
	\href{mailto:arjun.roy@bakerhughes.com}{\tt arjun.roy@bakerhughes.com}}\; and Joseph~P.~Cusumano\thanks{Corresponding author: \href{mailto:jpc3@psu.edu}{\tt jpc3@psu.edu}.}\\
    Department of Engineering Science \& Mechanics\\
	Penn State University\\
	University Park, Pennsylvania, USA
    }

\date{\today}

\begin{document}

\maketitle
\begin{abstract}
We study the non-linear dynamics and failure statistics of a coupled-field fatigue damage evolution model. We develop a methodology to derive averaged damage evolution rate laws from such models. We show that such rate laws reduce life-cycle simulation times by orders of magnitude and permit dynamical systems analysis of long-time behavior, including failure time statistics. We use the averaged damage rate laws to study 1 DOF and 2 DOF damage evolution models. We identify parameter regimes in which the systems behave like a brittle material and show that the relative variability for failure times is high for such cases. We also use the averaged rate laws to construct damage evolution phase portraits for the 2 DOF system and use insights derived from them to understand failure time and location statistics. We show that, for brittle materials, as the relative variability in failure time \textit{increases}, the variability in failure location \textit{decreases}.
\end{abstract}


\section{Introduction}

Fatigue failure is typically defined as the failure of specimens that have been subjected to cyclic loading \cite{Suresh2006}. Most industrial machinery and automated modes of transport, like aircraft and automobiles, have one or more internal components that are subject to cyclic loads and hence are prone to fatigue failure. However, despite many years of research devoted towards the development of  analytical models for fatigue (the earliest studies being done in the first half of the nineteenth century \cite{Suresh2006, Schutz1996}), a general understanding of failure time statistics in systems subject to fatigue remains elusive. This is because, first, experimental data on the life of specimens subjected to fatigue loading is difficult and expensive to obtain. Furthermore, such data is sensitive to minor differences arising, for example, from surface finish and microstructural variations or residual stresses and, thus, usually has a lot of variance. But, second, because fatigue is a complex process that involves the coupling of phenomena occurring at very different length and time scales, it is difficult to build mathematical models from first principles that can bridge the gap between these scales and still be computationally efficient, making it difficult to generate the large amounts of failure time data needed to study failure time statistics. In addition, the coupled field nature of damage physics has made it difficult to apply modern tools of dynamical analysis, particularly the state space methods of dynamical systems theory.

In this paper we use the method of averaging \cite{Sanders2007, Guckenheimer1983} to develop averaged damage \textit{rate laws}, from coupled-field damage evolution models. Such rate laws are computationally efficient and so can be readily used to study the system's nonlinear damage evolution and failure time statistics using numerical experiments and, at least in many cases, analytically. Using this approach, we aim to provide insight into the manner in which uncertainties, presumably arising from variations in internal microstructure, affect space-time failure statistics. We show how these averaged damage evolution laws allow us to derive approximate analytical results as different system properties are varied, thus opening up the potential to study damage rate laws with a degree of generality beyond what has previously been done.

The literature on modeling fatigue, or damage in general, has spread out in different directions. A significant portion of it deals with the mechanistic modeling of fatigue crack growth. This started with the work of Griffith \cite{GriffithA.1921} and Irwin \cite{Irwin1956, Sneddon1946, Westergaard1939, Irwin1957} focusing on deriving expressions for the stress intensity field near a crack-tip, which eventually led to the development of linear elastic fracture mechanics (LEFM) and elastic-plastic fracture mechanics \cite{Burdekin1966, Anderson1995, Rice1968, Rice1968a, Rice1968b, Hutchinson1968}. These ideas have been used to model fatigue crack growth \cite{Klesnil1972, Frost1958, Jones2008, Frost1971, Bilby1968} with rate laws resembling that suggested by Paris and Erdogan \cite{Paris1963}, and in which the effects of stress and geometry are accounted for via the stress-intensity factor. A limitation of such models is that they focus on modeling the growth of a single dominant crack in the continuum and therefore are not suitable for studying the evolution of multiple competitive microcracks or other micro-defects present before individual observable cracks have coalesced. 

An alternate approach for modeling damage evolution that aims to study changes at the  microstructural level caused, for example, by a large number of distributed micro-cracks oriented in different directions, is continuum damage mechanics (CDM). In CDM, field variables are first defined that quantify the average level of damage in continuum-scale volume elements of a material, then relations are derived that link these microstructural field variables to observable macroscopic field variables \cite{Krajcinovic1996, Kachanov1999, Rabotnov1963, Leckie1974, Leckie1977, F.A.Leckie1978, Murakami1980, Murakami1983, Budiansky1976}. The coupled field damage evolution model that we study in this paper falls under this class: a summary of the model is presented in section \ref{sec:Model}; a detailed derivation of the model can be found in \cite{Cusumano2015}. We assume here that the source of variability in failure time statistics lies in the uncertainty in the initial damage state---that is, in the microstructural damage field---and study the effect of different system parameters on the probabilistic evolution of the system of from such uncertain initial states. While the model studied here was chosen in the interest of specificity, the general approach we demonstrate can be expected to be applicable to a wide range of alternative formulations. 

The other direction in which the fatigue literature has evolved has centered around the use of a probabilistic framework, with the primary objective of estimating failure time statistics under different loading conditions typical in experiments \cite{VirklerD.A.1978, H.Ghonem1987}. The simplest approach involves estimating unknown parameters in one of the standard cumulative distribution functions (CDF) used to model failure time distributions \cite{NISTUrl} such as the log-normal distribution (\cite{Parzen1959,NISTUrl}) or the Birnbaum-Sanders distribution \cite{Desmond1985,Birnbaum1969,Birnbaum1958} or extreme-value distributions \cite{Galambos1987, Castillo2005,Galambos1987,Freudenthal1953,Weibull1951,Weibull1961}---each of which is selected based on some heuristic assumptions. Others treat damage as an abstract scalar random variable that evolves as per an evolution law. In some cases this evolution is modeled using discrete-state Markov chains with the transition probability matrix being estimated from experimental data \cite{Bogandoff1978, Bogandoff1978a, Bogandoff1978b, Bogandoff1980, Sobczyk1989, Ditlevsen1986a, Jones1973}. In other cases, damage is treated as a continuous time stochastic process and stochastic differential equations (SDEs) are used to model damage growth \cite{Oh1979,Oh1980,Lin1983,Dolinski1986,Ditlevsen1986,Sobczyk1986,Kozin1981}. However, in these cases the damage evolution rate law is not derived from general first principles; instead, it is usually a probabilistic version of the Paris-Erdogan crack growth law \cite{Paris1963}.

In this paper, a coupled-field damage evolution model (described in Section \ref{sec:Model}) is used to describe damage evolution under different loading conditions and for different values of intrinsic system parameters. We define damage, $\phi$, as a positive scalar micro-structural configuration variable, varying between 0 (undamaged) and 1 (failed), whose evolution is coupled with the dynamics of the macroscopic configuration variables (displacements) and whose initial state is uncertain and distributed uniformly over a small range ([0, $\phi_0$]). Critically, even though the subsystem governing macroscopic displacements for fixed damage is linear, the coupling between scales gives the system an overall nonlinear structure. To get failure time statistics, we perform ensembles of simulations that give us damage evolution trajectories (from the initial damage state until failure at $\phi = 1$) and, thereby, estimates of failure time distributions. However, even for the low-dimensional system studied here, this approach is computationally expensive. This is primarily because of the large disparity between the macroscopic and microscopic time scales, which requires that simulation time steps be small enough to capture the fast dynamics at the macroscopic scale, while  total life-cycle simulation times are long because damage evolves very slowly. 

We therefore turn to the method of averaging \cite{Sanders2007,Guckenheimer1983}: by averaging out the effect of the rapidly-oscillating macroscopic dynamics on the growth of damage at the microscopic scale, we thereby derive an averaged damage evolution rate law (Section \ref{sec:Averaging}). In previous work \cite{Cusumano2015}, averaging was applied to a limiting case of a 1D continuum version of the model studied here; it was shown to yield a rate law with the same form as the Paris-Erdogan model for crack growth. This correspondence was thereby used to put physically-meaningful bounds on damage model parameters. In this work, we show that using the averaged rate laws significantly reduces life-cycle computational time, potentially by many orders of magnitude, and permits analyses not otherwise possible. We first implement the method of averaging on a simple one degree of freedom (DOF) model that describes damage evolution in a spring-mass-damper system. We find a universal slow time scale over which damage growth is independent of the damage growth rate constant and the forcing amplitude. Further, the simplicity of this 1 DOF model helps us explore the effect of different parameters on the system. We identify parameter regimes in which the behavior of the system resembles that of brittle materials. We also derive analytical expressions for failure time CDFs and show the nature of $S$-$N$ curves that would be produced by our model (Section \ref{FTStats})

As a first step toward understanding the effect of system parameters on failure location (or failure \textit{mode}), as well as how these failure modes compete with each other, we next develop the averaged rate law for a 2 DOF spring-mass-damper damage model (Section \ref{2DOFdamage}). We are able to find a first integral for dynamics in the averaged \textit{damage phase space} and use it to construct phase portraits for different parameter regimes. Further, we demonstrate the existence of integral curves that partition the phase space into regions corresponding to different failure modes. Finally, we use the averaged damage evolution curves to study failure time and location statistics for the 2 DOF system (Section \ref{Failurestats}).


\section{Damage evolution model}
\label{sec:Model}

\begin{figure}[t] 
\begin{center}
\includegraphics[width=1\textwidth]{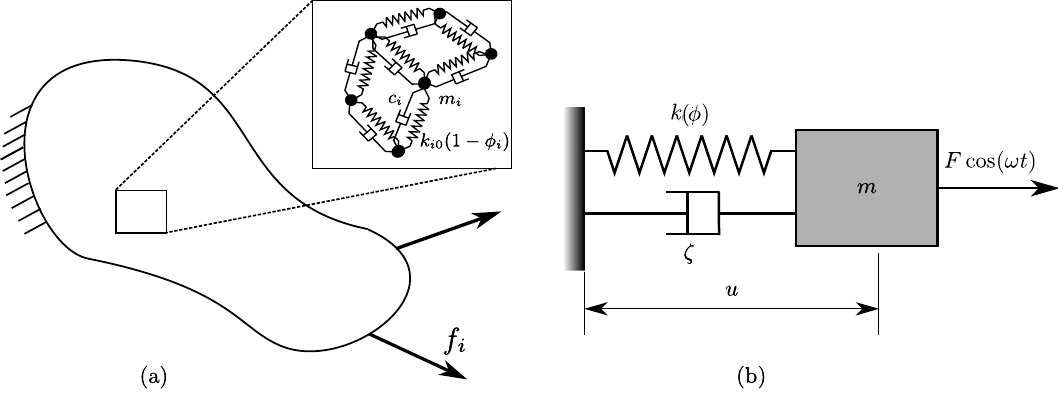}
\caption{(a) A spring-mass-damper model of a general body. (b) A 1 DOF spring-mass-damper model under periodic external load. In either case the stiffness of the springs alters as damage evolves}
\label{model}  
\end{center}
\end{figure}

We study the evolution of damage in a linear elastic body with internal structural damping under the action of external forces using the damage evolution law presented in \cite{Cusumano2015}. Damage is modeled as a microstructural scalar field that varies between 0 (undamaged) and  1 (fully-damaged) and is coupled to the macroscopic dynamics via the stored internal energy of the system. In this paper we consider a lumped mass approximation of the continuous system presented in \cite{Cusumano2015}, ie we approximate the body with a system of masses coupled by springs and dashpots, as shown in Fig.~\ref{model}(a). Such an approximation can potentially result from a finite-element model of the system. We assume that the external forces acting on this system are periodic as we intend to study the evolution of damage under fatigue loading and translate the essential and natural boundary conditions of the body to corresponding displacement constraints and periodic external point loads on the individual mass-points. 

In the present lumped mass approximation we therefore assume that the stiffness in (say), the $i$-th spring is dependent on damage in that spring, $\phi_i$, via a thermodynamically consistent relation, $k_i(\phi_i)$. Thus, damage in the system can be represented via an $r$-dimensional vector ${\boldsymbol{\upphi}}=[\phi_1,\phi_2,\ldots,\phi_r]^T$, whose individual components represent the damage level for each spring and where $r$ is (say) the number of springs in the system. Specifically, we assume that $k_i(\phi_i)= k_{i0}(1-\phi_i)$, where $k_{i0}$ is the undamaged stiffness. This definition of damage is similar to that presented in continuum damage mechanics literature \cite{Budiansky1976,Kachanov1986,Chaboche1988} and follows from the discussion presented in \cite{Cusumano2015}. However, the approach that we present here to derive an averaged damage evolution law is general and can be used for any given functional form of $k_i(\phi_i)$.

The equations of motion for this system thus have the following general form:
\begin{subequations}\label{ODEs}
\begin{align}
\mathsf{M}\ddot{\mathbf{u}}+\mathsf{C}\dot{\mathbf{u}}+\mathsf{K}(\boldsymbol{\upphi})\mathbf{u} = &\mathbf{f}e^{i\omega t} \label{FE_u}\\
\dot{\boldsymbol{\upphi}} = &\eta\mathbf{g}(\mathbf{u},\mathbf{\upphi})\label{FE_phi},
\end{align}
\end{subequations}
where $\mathbf u$ is a vector representing the macroscopic displacement of the masses, $\mathsf{M}$, $\mathsf{C}$ and $\mathsf{K}$ are the mass, damping and stiffness matrices, $\mathbf{f}$ and $\omega$ are the forcing amplitude and frequencies respectively and the overdots represent derivatives with respect to time. For the more general continuum case, the matrices $\mathsf{M}$, $\mathsf{C}$ and $\mathsf{K}$ could be generated via spatial discretization of the solution space using finite elements. The evolution of damage in this system is driven by the macroscopic strain energy of the system. This again follows from the discussion presented in \cite{Cusumano2015}, wherein Hamilton's Principle is used to derive damage evolution laws in which damage growth is driven by the strain energy dependent microforces. The elements of the vector $\mathbf{g}$ in Eq.~\eqref{FE_phi} gives the damage evolution law for damage in each spring and $\eta$ is the scaling parameter that controls the average rate at which damage evolves. Since, damage evolves in a time scale that is much smaller than the macroscopic dynamics, we assume that $0\le\eta\ll1$.

As in \cite{Cusumano2015}, we take initial damage state uncertainty to be the primary source of statistical variability observed in experimental failure time data. Therefore to study failure time statistics using this model we perform life-cycle simulations for an ensemble of random initial damage values to obtain failure time data. However, since damage evolves over a much slower time scale in comparison to the fast time scale of the macroscopic dynamics, the computational expense involved in simulating large ensembles of these models is prohibitive. We next use the method of averaging  \cite{Sanders2007} to derive an averaged damage evolution rate law that reduces the computational time required for these simulations by orders of magnitude.


\section{Damage evolution in a 1 DOF system}
\label{sec:Averaging}

We first derive the averaged damage evolution law on an equivalent 1 DOF system as seen in Fig.~\ref{model}(b). The equations of motion for this system are
\begin{subequations}\label{1-D-fully-coupled}
\begin{align}
m\ddot{u}+c \dot{u} +k\left(\phi\right) \,u & =  f\,\mathrm{cos}\left( \omega\,t\right) \label{macro1full}\\
\dot{\phi} & = \eta g(\phi,u), \label{micro1full}
\end{align}
\end{subequations}
where, $u$ is the displacement of the point mass, $m$, $c$ is the damping coefficient, $k(\phi)$ is the stiffness of the spring, $f$ is the forcing amplitude, $\omega$ is the forcing frequency and $\phi$ is the damage in the spring. The stiffness of the spring reduces as damage increases and is given by $k(\phi) = k_0\left(1-\phi \right)$, where $k_0$ is the undamaged stiffness of the spring. The evolution of damage is governed by Eq.~\eqref{micro1full}, where $\eta$ is the rate constant that controls damage evolution and $g(\phi,u)$ determines the damage evolution rate law. We substitute $\omega_n=\sqrt{k_0/m}$, $\zeta=c/2m\omega_n$ and $F=f/m$, in Eqs.~\ref{1-D-fully-coupled} to get:
\begin{subequations}\label{fully-coupled}
\begin{align}
\ddot{u}+2\,\zeta\,\omega_n \dot{u} +{\omega^2_n}\,\left( 1-\phi\right) \,u & =  F\,\mathrm{cos}\left( \omega\,t\right) \label{macro1}\\
\dot{\phi} & = \eta g(\phi,u), \label{micro1}
\end{align}
\end{subequations}
where, $\omega_n=\sqrt{k_0/m}$ is the undamaged natural frequency of the system, $\zeta=c/2m\omega_n$ is the damping factor and $F$ is the forcing amplitude. 
When $\eta=0$, there is no damage evolution and the system behaves linearly with a steady state (for any arbitrary damage value $\phi$) given by:
\begin{subequations}\label{ss_both}
\begin{equation}\label{ss_sol}
u_{ss}(t;\phi) = F A(\phi) \cos(\omega t + \theta),
\end{equation}
\begin{equation}\label{A_and_theta}
\mbox{where},\quad A(\phi)=\dfrac{1}{ \sqrt{ \left [ \left (1 - \phi\right)\omega_n^2 -\omega^2 \right ]^2 +\left ( 2\zeta \omega\omega_n \right )^2 }}\quad \mbox{and}\quad \theta = \tan^{-1}\left (\dfrac{2\zeta \omega \omega_n}{(1 - \phi)\omega_n^2 -\omega^2}\right )
\end{equation}
\end{subequations}

As mentioned in the previous section, because damage evolves at a much slower time scale as compared to the macroscopic dynamics, $0<\eta\ll 1$. We thus set $\eta$ to be the smallness parameter in our subsequent calculations. This further implies that the transients induced due to small changes in $\phi$ are short-lived in comparison to the time scale over which $\phi$ evolves. Thus as damage evolves, the macroscopic system stays within $\mathcal{O}(\eta)$ of its unperturbed ($\eta=0$) steady-state. Hence, in such a scenario we can say that,    
\begin{equation}\label{macro_u}
u(t)=u_{ss}(t;\phi)+ \mathcal{O}(\eta).
\end{equation}

On substituting Eq.~\eqref{macro_u} and~\eqref{ss_sol} in Eq.~\eqref{micro1}, we  get an equation which is in the `standard form' that is necessary for applying the method of averaging \cite{Sanders2007,Guckenheimer1983}. We therefore obtain an averaged damage evolution law as follows:
\begin{align}
\dot{\phi}  &=  \frac{\omega}{2\pi}\int_0^{\dfrac{2\pi}{\omega}}\eta g\left (\phi,u_{ss}(x; \phi) \right) dx \label{microavgthresh} 
\end{align}
Solutions to Eq.~\eqref{microavgthresh} will stay within $\mathcal{O}(\eta)$ of solutions to Eq.~\eqref{micro1} on a time scale $t \sim \mathcal{O}(1/\eta)$ \cite{Guckenheimer1983} for any given form of $g(\phi,u)$. Specifically, in this paper we use a form of $g(\phi,u)$ that has been derived previously in \cite{Cusumano2010,Cusumano2015} using Hamilton's Principle and has the following form:
\begin{equation}
\label{dam_growth}
g(\phi,u) = \eta \phi^{-q/p} \left (\left [ \frac{{u}^{2}}{2} - \frac{\alpha}{\phi^{1/3}}\right ]^{+}\right )^{1/p}.
\end{equation}
Here the\ `$+$' superscript denotes the positive part of the quantity within the brackets(i.e. $[b]^+= (b+|b|)/2, \forall b\in \mathbb{R}$) . Damage growth in the spring is driven by the strain energy in the spring (proportional to $u^2/2$) and is positive as long as $u^2/2$ exceeds the threshold $\alpha\phi^{-1/3}$. The intensity of this threshold is controlled by the parameter $\alpha$ and parameters $p$ and $q$ yield exponents that control the average damage growth exponent. It can be shown \cite{Cusumano2010,Cusumano2015} that a limiting case of this model is the well-known Paris's rate law \cite{Paris1963,Suresh2006} and that for metals the exponents $p$ and $q$ are expected to lie in the ranges: $1/2\le p \le 1$ and $-1\le q \le -2/3$ and thus $-2\le q/p\le-2/3$. Substituting Eqs.~\ref{dam_growth} and \ref{ss_sol} in Eq.~\ref{microavgthresh}, we get
\begin{equation}
\label{avg_dam_sp_form}
\dot{\phi}  =  \frac{\omega}{2\pi}\int_0^{\dfrac{2\pi}{\omega}}\eta\,\phi^{-q/p}\,\left ( \left [ \frac{1}{2}F^2 A(\phi)^2 \cos^2(\omega x + \theta) - \frac{\alpha}{\phi^{1/3}}\right ]^{+}\right )^{1/p} dx,
\end{equation}

It is clear from Eq.~\eqref{avg_dam_sp_form} that for forcing amplitudes $F\le F_E(\phi)\triangleq2\alpha/\left ( A(\phi)^2\phi^{1/3}\right )$, there will be no damage growth. Thus $F_E(\phi)$ serves as the endurance limit \cite{Suresh2006} for our model and we can see that it reduces as damage increases.
Now, if we rescale time as $\tau = \eta t$, then we get a rate law which is independent of $\eta$ as follows:
\begin{equation}
\label{avg_dam_no_eta}
\dot{\phi}  =  \frac{\omega}{2\pi}\int_0^{\dfrac{2\pi}{\omega}}\,\phi^{-q/p}\,\left ( \left [ \frac{1}{2}F^2 A(\phi)^2 \cos^2(\omega x + \theta) - \frac{\alpha}{\phi^{1/3}}\right ]^{+}\right )^{1/p} dx,
\end{equation}
The overdot in Eq.~\eqref{avg_dam_no_eta}, thus now represents derivatives with respect to $\tau$. Note, that this rescaling essentially corresponds to shifting to the slower time scale at which damage evolves. It helps in reducing computational time  involved in performing life-time simulations significantly as we can now march along the slower time scale $\tau$. Thus, the time required for integrating the rescaled damage rate law does not change as $\eta$ is reduced and hence the speed-up we get by using the rescaled average damage rate laws increases as $\eta$ is reduced.

In the absence of a threshold (i.e. for $\alpha=0$), we get a closed-form of the damage evolution law as follows:
\begin{subequations}\label{phidot_and_ab}
\begin{equation}\label{phidotsimp}
\dot{\phi}= \frac{ a_1\phi^{-q/p}}{ \left ( a_2 + a_3\phi +a_4\phi^2 \right ) ^{1/p}}
\end{equation}
\begin{align}
\mbox{where}, 
 \quad a_1 = \displaystyle\frac{\,F^{2/p} }{2^{1+1/p} \pi}  \int_0^{2\pi}\,\left [ \cos^2(y )\right ]^{+1/p} dy, \quad &a_2=\omega_n^4+4\zeta^2\omega^2\omega_n^2+\omega^4-2\omega^2\omega_n^2, \notag\\
&a_3=-2\omega_n^2\left ( \omega_n^2- \omega^2\right )\quad \mbox{and}\quad a_4=\omega_n^4\label{ab}
\end{align}
\end{subequations}

\begin{figure}[t] 
\begin{center}
\includegraphics[width=1\textwidth]{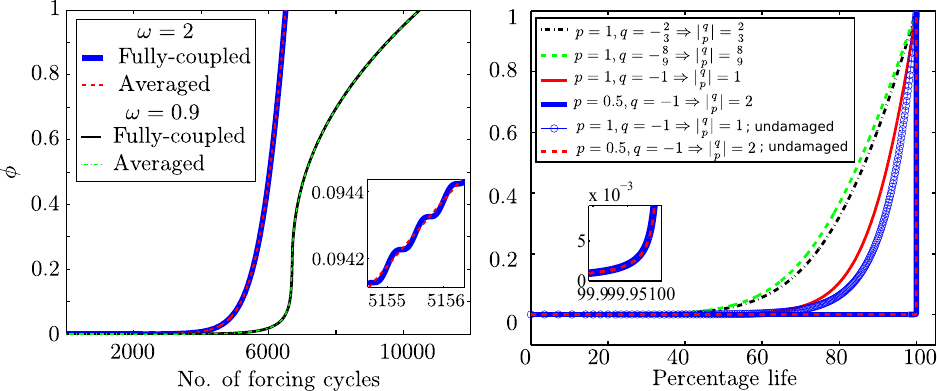}
\caption{\emph{(left-hand side plot)} Comparison of damage evolution curves obtained by integrating the averaged and the fully-coupled system. (\emph{curve on left}) Parameter values: $\omega = 1$, $\omega_n = 0.5$, $\eta = 0.0001$, $\zeta = 0.003$, $F = 2.5$ and $\phi_0 = 1 \times 10^{-6}$. Time-step size for either simulation : $\Delta t = \Delta \tau = 5\times 10^{-4}$. (curve on \emph{right}) A different damage evolution curve that is obtained when there is an early passage through resonance. Parameter values: $\omega = 1$, $\omega_n = 1.1$, $\eta = 0.0001$, $\zeta = 0.1$, $F = 3.0$ and $\phi_0 = 1\times 10^{-6}$. \emph{(right-hand side plot)} Damage evolution curves for different values of $p$ and $q$. Parameter values: $\omega = 1$, $\omega_n = 0.5$, $\eta = 0.0001$, $\zeta = 0.003$, $F = 2.5$ and $\phi_0 = 1 \times 10^{-6}$ }
\label{avgfull_subres}  
\end{center}
\end{figure}
Fig.~\ref{avgfull_subres} {\emph{(left)} shows a comparison of damage evolution trajectories obtained by integrating Eqs.~\eqref{fully-coupled} and \eqref{phidotsimp} for different parameter values. We see that the averaged damage evolution trajectory follows the fully-coupled trajectory up to failure. The inset further shows that the averaged damage evolution trajectory follows the mean of the oscillating damage evolution trajectory obtained using the fully-coupled system. Both the simulations were performed in Matlab using its inbuilt {\tt ode45} function with the same accuracy settings (viz. parameters {\tt abstol} =$1\times10^{-8}$ and {\tt reltol} = $1\times10^{-8}$). The initial conditions of the macroscopic system in Eq.~\eqref{fully-coupled} corresponded to its steady state initial conditions based on the initial damage state. The time required to integrate the rescaled averaged equation is very small in comparison to the time taken by the fully-coupled system. For example, for\  $F=16,\ \omega=2,\ \omega_n=1,\ \phi(0)=1\times10^{-6},\ \zeta=0.003,\ \alpha=0,\ p=1$ and $q=-1$ the time required to integrate Eqs.~\eqref{fully-coupled} is 7000 times than that taken by Eq.~\eqref{phidotsimp}. Further, since the rescaled average damage evolution is independent of $\eta$, the speed-up obtained increases as $\eta$ is reduced (i.e. damage evolution rate is made slower and slower). Given the very small value of $\eta$ expected in realistic applications, particularly for high cycle fatigue, we can expect the averaged equation to reduce life-cycle calculation times by multiple orders of magnitude thereby expediting the calculation of failure time statistics.

Also, note that as damage evolves the natural frequency of the system reduces and hence if $\omega$ is less than the undamaged natural frequency $\omega_n$, then the system passes through resonance when $\omega_n\sqrt{1-\phi}=\omega$. In particular, if this occurs early in the life of the specimen then the amplitude of the macroscopic displacement (and thereby the potential energy of the spring that drives damage growth) increases suddenly by a large amount and this then leads to a sudden increase in damage. Thus, in such a scenario, we obtain damage evolution curves that have a point of inflexion as shown in Fig.~\ref{avgfull_subres} ({\emph{left}).

We also investigate the effect of the exponents $p$ and $q$ on the nature of damage evolution curves. For this we perform simulations for different values of $p$ and $q$ holding other parameters constant.
Fig.~\ref{avgfull_subres} (\emph{right}) shows plots of damage as a function of percent life spent for different values of $p$ and $q$. We first note that damage evolution curves for different combinations of $p$ and $q$, with the same $\vert q/p\vert$ ratio are also close --- thereby indicating that the ratio $\vert q/p\vert$ primarily determines the exponent of the damage evolution power law. We also see that as the ratio $\vert q/p\vert$ increases, the growth of damage tends to become more sudden and concentrated towards the end of life of the specimen. This therefore indicates that on increasing $\vert q/p\vert$ the damage model tends to behave more like a brittle material. Furthermore, as shown in \cite{Cusumano2015}, the parameters $p$ and $q$ can be related back to the material-dependent Paris' law exponent, $m$ \cite{Suresh2006}. Using this relation we can conclude that our model suggests that the Paris' law exponent would be higher for brittle materials. This fact matches experimental results found in the literature, wherein it has been reported that the values for $m$ (obtained from experiments) are much higher for brittle materials ($m\sim 15\--50$), such as ceramics, than for ductile materials ($m\sim 2\--4$ for metals) \cite{Ritchie1999,Ritchie2000,Dauskardt1992,Gilbert1996,Liu1991}.

Further, since damage stays close to zero for large values of $\vert q/p\vert$, the unperturbed macroscopic steady-state, $u_{ss}(t,\phi)$ also stays close to the \emph{undamaged} macroscopic steady-state $u_{ss}(t,\phi=0)$.  
If we now substitute the undamaged macroscopic steady-state solution in Eq.~\eqref{dam_growth}, and following the same procedure to obtain the rescaled damage evolution law, we obtain a simpler damage rate law (for $\alpha=0$) as follows:
\begin{equation} 
\label{undam}
\dot \phi = \frac{a_1\phi^{-q/p}}{a_2^{1/p}}.
\end{equation}
In Fig.~\ref{avgfull_subres} (\emph{right}), we see that the damage growth curves obtained using Eqs.~\eqref{undam} and ~\eqref{avg_dam_sp_form} match very well in the when $|q/p| \rightarrow 2$, but not so well when away from it (eg. when $|q/p| = 1$). Also, the error in estimates of failure time are significantly higher when $|q/p| = 1$ (49.6\%) as compared to that when $|q/p| = 2$ (0.002\%). Thus, although the undamaged macroscopic steady state results in a simpler averaged damage rate law, it can be used only in the limit $|q/p| \rightarrow 2$, wherein damage growth is negligible for most of the life of the specimen.


\section{Failure time statistics}
\label{FTStats}

We next use the averaged damage rate law to study failure time statistics. The analysis presented here starts from the more general analysis presented in \cite{Cusumano2000}, with a focus on our present damage evolution model.  We assume that the random initial damage state is uniformly distributed over a small positive interval, say $[0,\;\phi_{\max}]$ and use this to calculate approximate analytical failure time distributions.

The failure time, $T$, of the rescaled averaged system (for $\alpha = 0$) starting from an initial damage state $\phi_0$ can be calculated using Eq.~\eqref{phidotsimp} as
\begin{align}
T&=\int_{\phi_0}^1 \frac{1}{a_1\phi^{-q/p}}\left( a_2 + a_3\phi +a_4\phi^2 \right )^{1/p} d\phi \notag \\
&=\int_{\phi_0}^1 \frac{a_2^{1/p}}{a_1\phi^{-q/p}}\left( 1 + d_1\phi +d_2\phi^2 + \ldots \right )d\phi, \\
&=\begin{cases}
\displaystyle\frac{a_2^{1/p}}{a_1\left(1+q/p\right)} \left ( 1-\phi_0^{1+ q/p}\right )+ \displaystyle\frac{a_2^{1/p}}{a_1}
\displaystyle\sum_{i=1}^\infty \frac{d_i}{i+1+q/p}\left (1-\phi_0^{i+1+ q/p}\right)\;&(q/p\ne-1)\\
\label{T_series}\\
-\displaystyle\frac{a_2^{1/p}}{a_1}\ln \phi_0 
+ \displaystyle\frac{a_2^{1/p}}{a_1}
\displaystyle\sum_{i=1}^\infty \frac{d_i}{i}\left (1-\phi_0^{i}\right)\qquad \; &(q/p=-1)
\end{cases}
\end{align}
where, $\phi_0$ is the initial damage and the coefficients $d_i$ result from a generalized binomial series expansion as
\begin{equation}
\label{di}
d_i=\sum_{k=0}^i \frac{1/p(1/p-1)\ldots(1/p-k+1)}{k!}\binom{k}{i-k}\left (\frac{a_3}{a_2}\right )^{2k-i}\left (\frac{a_4}{a_3}\right )^{i-k}, \qquad i=1\ldots\infty.
\end{equation}

We make use of the fact that $0<\phi_0\le\phi_{\max}\ll1$ and approximate $T$ to $\mathcal{O}(1)$. The closed form expression for this approximation depends on parameters $p$ and $q$ and is as follows:
\begin{subnumcases}{T = }
\frac{a_2^{1/p}}{a_1\left ( 1 + q/p\right ) } + \frac{a_2^{1/p}}{a_1}
\sum_{i=1}^\infty \frac{d_i}{i+1+q/p}+\mathcal{O}(\phi_0^{1+q/p}) \qquad \;&( $-1 < q/p \le -2/3$) \label{qbyp_less_than_2by3}\\
-\frac{a_2 ^{1/p}}{a_1}\ln \phi_0
+ \displaystyle\frac{a_2^{1/p}}{a_1}
\displaystyle\sum_{i=1}^\infty \frac{d_i}{i}\ +\mathcal{O}(\phi_0) \qquad \; &( $q/p=-1$)\label{qbyp_1}\\
K_1 \left ( \frac{1}{\phi_0^{- q/p -1}}\right ) + \left ( K_2 - K_1 \right )+ \mathcal{O}(\phi_0^{2+q/p})\qquad \;&($-2 < q/p < -1$), \label{qbyp_brittle} 
\end{subnumcases}
where, $K_1=\frac{a_2^{1/p}}{a_1\left(-q/p -1\right)} $ and $K_2=-\frac{a_2^{1/p}}{a_1\left(-q/p -1\right)} + \frac{a_2^{1/p}}{a_1}\sum_{i=1}^\infty \frac{d_i}{i+1+q/p}$. 

We see that for $-1< q/p \le -2/3$, $T$ is a constant to $\mathcal{O}(1)$ and hence the relative variability of $T$ in this case will be low as the only source of randomness in our simulations is the initial damage value.
For $q/p=-1$, if we use the $\mathcal{O}(1)$ approximation of $T$ and assume that $\phi_0$ is uniformly distributed over the interval $[0,\phi_{\max}]$, we can get a closed form approximation of the failure time cumulative distribution function ($F_{T}$) as follows:
\begin{equation}
   F_T(t) = \left\{
     \begin{array}{lr}
0 &\; (t < T_{\min})\\
       1 - \exp\left ( -\displaystyle\frac{a_1}{a_2^{1/p}} \left( t-T_{\min}\right )\right ) & \;(t \ge T_{\min})
     \end{array}
   \right.,
 \label{T_cdf}
\end{equation}
where, $T_{\min}=-a_2^{1/p}/a_1\ln(\phi_{\max})+a_2^{1/p}/a_1\sum_{i=1}^\infty \frac{d_i}{i}$, is the failure time corresponding to the initial condition $\phi_0=\phi_{max}$ and hence is the shortest possible failure time for the given ensemble. Thus we see that in the absence of a threshold and for $q/p=-1$, the failure time distribution is exponential in nature. 
\begin{figure}[t]
\begin{center}
\includegraphics[width=\textwidth]{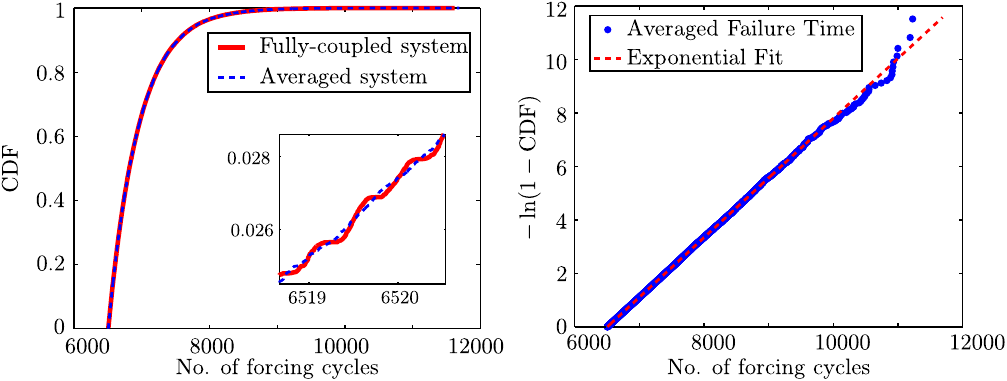}
\caption[Comparison of failure time distributions obtained using the averaged and fully-coupled system]{(\emph{left})Comparison of failure time distributions obtained using the averaged and fully-coupled system. Parameter values: $F =6$, $\omega = 1$, $\omega_n = 0.5$, $\eta = 0.0001$, $p=1$, $q=-1$ and $\zeta = 0.003$. $\phi_0$ is uniformly distributed on $[0,\;1\times 10^{-6}]$(\emph{right}) CDF for same parameters plotted on an exponential plot.}
\label{CDF}  
\end{center}
\end{figure}

In Fig.~\ref{CDF}, we present a comparison of failure time statistics, obtained by calculating the failure time, using both the averaged and fully-coupled system, when $p=1$ and $q=-1$. For this we performed life-cycle simulations for an ensemble of oscillators starting from random initial damage values that were uniformly distributed over $[0,1\!\times\!10^{-6}]$. We see that the match between the probability distributions of the failure time obtained using the averaged and fully-coupled solution is very good. In fact the maximum absolute difference in the calculated failure times over the entire ensemble is less than one-fourth of a forcing period. Further, the Kolmogorov-Smirnov hypothesis test (by using Matlab's {\tt kstest2} function) also indicates that the two distributions are close. The inset in Fig.~\ref{CDF} shows that while the failure time distribution obtained using the fully-coupled system has oscillations in it (due to the influence of the macroscopic dynamics), the distribution obtained using the averaged equations does not. Further, we also see that the periodicity of these oscillations is twice that of the forcing period. This is due to the fact that our damage law allows for damage to grow both in the tension and compression halves of the forcing cycle \cite{Cusumano2015}. A plot of the empirical CDF obtained using the averaged damage evolution law on an exponential probability paper plot, as seen in Fig.~\ref{CDF} (\emph{right}), demonstrates that the failure time CDF is exponential in nature as shown by Eq.~\eqref{T_cdf}.

For $-2\le q/p<-1$, as $q/p\rightarrow-2$, the leading term in Eq.~\eqref{qbyp_brittle} will approximately be $\mathcal{O}(1/\phi_0)$ and will be much larger than that of the $\mathcal{O}(1)$ term. The approximate failure time distribution using the leading term in Eq.~\eqref{qbyp_brittle} is as follows: 
\begin{equation}
   F_{T}(t) = \left\{
     \begin{array}{lr}
       0 &\; \left(t < \frac{K_1}{\phi_{\max}^{-1-q/p}}\right)\\
       1 - \dfrac{1}{\phi_{\max}}\left (\dfrac{K_1}{ t}\right )^{-1/(1+q/p)} & \;\left(t \ge \frac{K_1}{\phi_{\max}^{-1-q/p}}\right)
     \end{array}
   \right..
   \label{CDFTapp_ud}
\end{equation}

Also, as mentioned in the previous section, as $q/p\rightarrow-2$ the damage evolution resembles that observed in brittle fatigue and we can thus use Eq.~\eqref{undam} to calculate failure time. On using this approximation, we get the same expression for the failure time CDF as in Eq.~\eqref{CDFTapp_ud}.
Thus as $q/p\rightarrow-2$, the nature of the failure time distribution tends to a Pareto distribution \cite{Paretolink}. In Fig.~\ref{error_CDF} we see a comparison of numerically obtained failure time distributions with the analytical approximation Eq.~\eqref{CDFTapp_ud} for different values of $q/p$ ($q$ is held constant at -1, while $p$ is reduced from 0.95 to 0.5). We see that when $q/p$ is very close to -1, the approximation given by Eq.~\eqref{CDFTapp_ud} differs from the empirical CDF by just a constant and that the approximation converges to approaches the empirical CDF as soon as $p<0.87$. Further the nature of the distribution also deviates away from an exponential distribution, as shown in Eq.~\eqref{CDFTapp_ud}.

\begin{figure}[t]
\begin{center}
\includegraphics[width=\textwidth]{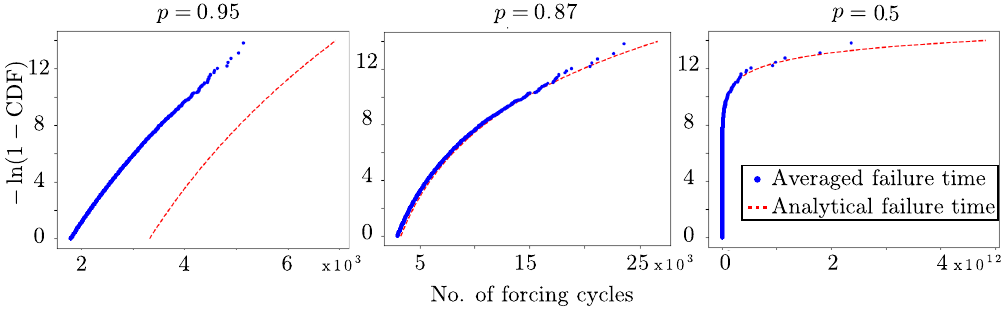}
\caption[Exponential plot of failure time distributions for different values of $p$]{Exponential plot of failure time CDFs for different values of $p$. Other parameter values same as in Fig. \ref{CDF}}
\label{error_CDF}  
\end{center}
\end{figure}

Next, to estimate the expected amount of `scatter' in failure times we define the relative variability $R$ as:
\begin{equation}
\label{R_orig}
R= \frac{T_{[95]}-T_{[5]}}{T_{[50]}},
\end{equation}
where $T_{[r]}$ is the $r$-th percentile of $T$. The expressions for $R$ corresponding to the failure time CDFs given by Eqs.~\eqref{T_cdf} and ~\eqref{CDFTapp_ud} are as follows:
\begin{equation}
\label{R}
R=\begin{cases}
\dfrac{a_2/a_1 \ln19}{T_{\min}+a_2/a_1\ln2}  \qquad \;&(q/p=-1; \mbox{ using Eq}.~\ref{T_cdf}) \\ \\
10^{-1-q/p}\left(1-\dfrac{1}{19^{-1-q/p}}\right) \qquad \;&(q/p\rightarrow-2; \mbox{ using Eq}.~\ref{CDFTapp_ud}) 
\end{cases}
\end{equation}

We see that for $q/p\rightarrow-2$, $R$ depends only on the damage rate law exponent $q/p$. The contour plot in Fig.~\ref{R_FN}(\emph{left}) shows how $R$ varies as a function of $|q/p|$ and $\omega$. We see that for most values of $|q/p|$, $R$ is independent of $\omega$ and is a constant except when it is close to resonance. This is because close to resonance the $\mathcal{O}(1)$ and $\mathcal{O}(1/\phi_0)$ approximations to $T$ do not hold good as the coefficients of the higher order terms in the expansion of $T$ become significantly large. Also, we see that $R$ increases in value as $q/p\rightarrow-2$. Thus, the sensitivity of the failure time to uncertainties in the initial damage value increases as $|q/p|$ increases (ie as damage evolution resembles brittle fatigue).

\begin{figure}[t] 
\begin{center}
\includegraphics[width=\textwidth]{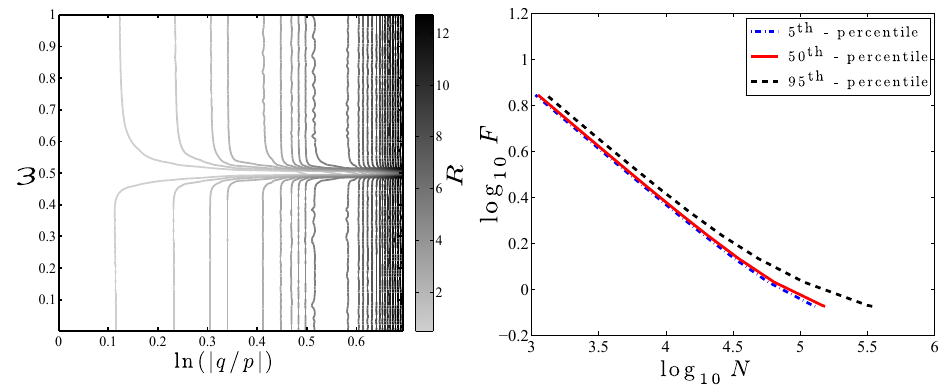}
\caption[Plots of variation of relative temporal variability with forcing frequency and $r$]{\emph{(left)} Pseudocolor plot showing the variation of relative temporal variability of the failure time with $p$ and $\omega$.\emph{(right)} $F$-$N$ curve. Parameter values: $\omega_n=0.5,\, \zeta=3\times 10^{-3},\, \hat{q}=1,\, \phi_0\sim$\,\, Unif\, $[0\,,1\times 10^{-6}]$, No.of samples = $1\times 10^6$,\, $\omega=1$. For the right plot, $\alpha = 5\times10^{-3}$, other parameters are the same as in the left plot.}
\label{R_FN}  
\end{center}
\end{figure}

We next use the averaged equations to study the nature of $F$-$N_f$ curves for the 1 DOF damage model, where $N_f$ is the number of forcing cycles to failure (equivalent to the $S$-$N$ curves that we typically see in the literature \cite{Suresh2006}). Using Eq.~\eqref{T_series}, we note that, in the absence of a threshold, we can express the failure time, $T$, as:
\begin{equation}
T = \frac{C_1\left ( \omega_n, \zeta, \omega, \phi_0, p, q \right)}{a_1} = \frac{C_2\left ( \omega_n, \zeta, \omega, \phi_0, p, q \right)}{F^{2/p}}
\end{equation}
Thus by rescaling failure time to load-cycle units, taking logarithms on both sides of this relation and rearranging terms, we obtain 
\begin{equation}
\ln (F)\,= -\frac{p}{2} \ln (N_f) + C,
\label{logFlogN}
\end{equation}
where, $C = p/2\ln \left(C_2\omega/\left(2\pi\right) \right)$. Thus, in the absence of a threshold, the $F$-$N_f$ curves, plotted on a log-log scale are always straight lines with a slope of $-p/2$. Further, since the source of variability of failure time is in the initial damage state, the statistical `spread' in the $\ln F$-$\ln N_f$ would be identical at all forcing levels. 
Even in the presence of a damage threshold ($\alpha \ne 0$), this holds true as long as the forcing amplitude is much larger than endurance limit (thereby ensuring that damage grows during most part of each forcing cycle). 
 This can be seen in Fig.~\ref{R_FN} (\emph{right}), wherein we see that the distance between the $5^{\mbox{th}}$ and $95^{\mbox{th}}$ percentile curves stays constant for larger values of $F$. As we approach the endurance limit, we find that the slope of the $F$-$N_f$ becomes more negative and that the statistical spread in the failure times increases as $F$ is reduced.


\section{Averaged damage phase space}
\label{2DOFdamage}

We next use the method of averaging to simulate damage evolution in a 2 DOF system. We present a generic approach to derive the averaged damage evolution laws for any 2 DOF system. However, for the ease of discussion and demonstrating specific results, we will then consider a specific 2 DOF spring-mass-damper system, forced at one end and fixed at the other, as shown in Fig.~\ref{2dspring_mass}. In this case damage evolves simultaneously in the two springs and thus we now have two scalar damage variables $\phi_1$ and $\phi_2$. The rate at which damage evolves in either spring is, as in the previous section, proportional to the strain energy of the corresponding spring. The macroscopic coupling between the spring-mass-damper systems couples the damage evolution at the microscopic level and in the ensuing sections we discuss the effects of different macroscopic parameters on damage evolution and failure statistics. Also, as in the previous sections, damage varies over the range $[0, 1]$ and the failure of the system is now defined as the moment when damage in any one of the springs reaches $1$. Thus the  of the system is the time when $\max (\phi_1, \phi_2)$ reaches 1 and the failure mode is said to be 1 if $\phi_1$ reaches 1 earlier and 2 if $\phi_2$ reaches 1 earlier.

The equations of motion for the macroscopic system are given by
\begin{figure}[t]
\begin{center}
\includegraphics[width=0.65\textwidth]{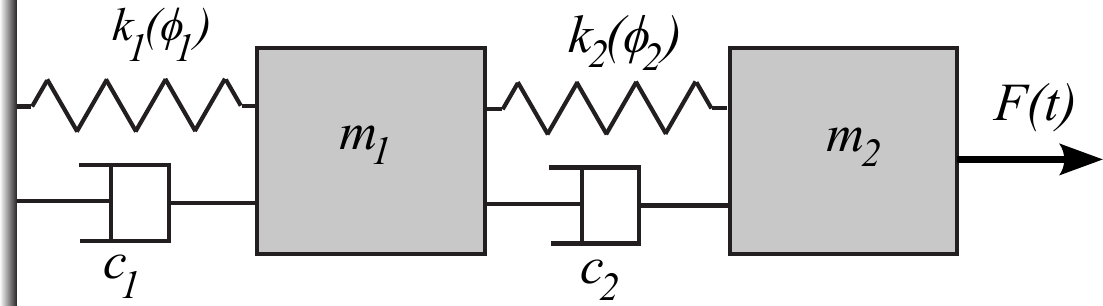}
\caption{A two-dimensional coupled spring mass model with damage}
\label{2dspring_mass}  
\end{center}
\end{figure}

\begin{equation}
\textbf{M}\ddot{\textbf{u}}+\textbf{C}\dot{\textbf{u}}+\textbf{K}\textbf{u}=\textbf{f}e^{i\omega t}\label{matrixeqn},\\
\end{equation}
where, $\textbf{M}$, $\textbf{C}$ and $\textbf{K}(\phi_1,\phi_2)$ are the mass, damping and stiffness matrices, $\textbf{u}$ is the displacement vector, $\textbf{f}$ is the forcing vector and $\omega$ is the forcing frequency. The stiffness matrix is a function of the scalar damage variables $\phi_1$ and $\phi_2$. The damage evolution laws for this system are given by
\begin{subequations}\label{damageevol}
\begin{align}
\dot{\phi}_1=&\  \eta{\phi}_{1}^{-q/p}\left [ \frac{|V_{1p}e^{i\omega t}|^2}{2}-\frac{\alpha}{\phi_1^{1/3}}\right ]^{+1/p} \label{2dphi1} \\
\dot{\phi}_2=&\  \eta{\phi}_{2}^{-q/p} \left[\frac{|V_{2p}e^{i\omega t}|^2}{2} - \frac{\alpha}{\phi_2^{1/3}} \right]^{+1/p},\label{2dphi2}
\end{align}
\end{subequations}
where, $V_{1p}$ and $V_{2p}$ are the amplitudes of the strains in spring 1 and 2 respectively and are a function of $\phi_1$ or $\phi_2$. As in the previous section, the damage growth is proportional to the strain energy in either spring.
To calculate the averaged damage evolution equations we proceed as in section~\ref{sec:Averaging} and assume that $0<\eta \ll 1$ and that the damping factor is sufficiently large to ensure that the transients in the macroscopic system are short-lived. These assumptions allow us to further assume that the macroscopic system always stays within $\mathcal{O}(\eta)$ of its unperturbed ($\eta=0$) steady-state. In other words, 
\begin{subequations}\label{2DSSeqn}
\begin{align}
u_1(t) & = u_{ss1}(t)+ \mathcal{O}(\eta)\label{2DSSeqn1}\\
u_2(t) & = u_{ss2}(t)+ \mathcal{O}(\eta)\label{2DSSeqn2}.
\end{align}
\end{subequations}
We thus next calculate the unperturbed steady state solutions to the macroscopic system, substitute them into the damage evolution law and average them over one forcing period to get the following average damage evolution laws:
\begin{subequations}\label{averageint}
\begin{align}
\dot{\phi}_1=&\  \frac{\omega}{2\pi}\int_0^{\dfrac{2\pi}{\omega}}  \eta{\phi}_{1}^{-q/p}\left [ \frac{|V_1e^{i\omega t}|^2}{2}-\frac{\alpha}{\phi_1^{1/3}}\right ]^{+1/p} dt\label{2davgint1} \\
\dot{\phi}_2=&\  \frac{\omega}{2\pi}\int_0^{\dfrac{2\pi}{\omega}}  \eta{\phi}_{2}^{-q/p} \left [\frac{|V_2e^{i\omega t}|^2}{2} - \frac{\alpha}{\phi_2^{1/3}} \right ]^{+1/p} dt,\label{2davgint2}
\end{align}
\end{subequations}
where $V_1$ and $V_2$, now represent the unperturbed macroscopic strain in spring 1 and 2 respectively. For the case where $\alpha =0$, we can rescale time as follows to simplify the averaged damage evolution laws
\begin{equation}\label{tscale_2DOF}
\tau =\left (\frac{\eta f^{2/p}}{2^{1+1/p}\pi} \int_0^{2\pi} \left ( \cos^2(y)\right )^{1/p} dy \right )t.
\end{equation}
Using the substitution in Eq.~\ref{tscale_2DOF}, we get the following rescaled averaged damage evolution equations:
\begin{subequations}\label{rescavgalpha0_2DOF}
\begin{align}
\dot{\phi}_1=&\phi_1^{-q/p}|V_{1}(\phi_1,\phi_2)|^{2/p}\label{avgalpha01_2DOF}\\
\dot{\phi}_2=& \phi_2^{-q/p}|V_{2}(\phi_1,\phi_2)|^{2/p}\label{avgalpha02_2DOF},
\end{align}
\end{subequations}
where the overdot now represents derivatives with respect to $\tau$. Thus, we see that the averaging approach discussed in section~\ref{sec:Averaging} is generic and can be easily extended to higher dimensions.

In order to discuss properties of this 2 DOF system, we further assume that $u_1$ and $u_2$ are the displacements of the two masses, $m_1$ and $m_2$ respectively, as shown in Fig.~\ref{2dspring_mass} and $\mu$ is the ratio of the two masses, such that $m_1 = \mu m_2$. The stiffness of the springs ($k_1$ and $k_2$) are assumed to depend on the respective damage variables as $k_1 = \omega_n^2 (1-\phi_1)m_1$ and $k_2 = \omega_n^2 (1-\phi_2)m_2$, the damping in the two dampers are assumed to be $c_1 = 2\zeta\omega_n m_1$ and $c_2 = 2\zeta\omega_n m_2$ and the magnitude of the external force acting on $m_2$ is assumed to be $F$. These assumptions result in the following form for the mass, stiffness and damping matrices and the force and displacement vectors:

\begin{equation}
\textbf{M} = \left( \begin{array}{cc}
1 & 0 \\
0 & 1  \end{array} \right)\quad \textbf{C} = \left( \begin{array}{cc}
2\zeta \omega_n(1+\mu) & -2 \zeta\mu \omega_n \\
-2 \zeta \omega_n  & 2 \zeta\omega_n   \end{array} \right )\quad	 \textbf{u} = \left( \begin{array}{c}
u_1  \\ u_2  \end{array} \right ),$$
$$ \textbf{K} = \Bigg( \begin{array}{cc}
\omega_n^2 ( 1 - \phi_1) + \mu\omega_n^2 (1 - \phi_2) & -\omega_n^2 \mu(1 - \phi_2) \\
-\omega_n^2 (1 - \phi_2) & \omega_n^2  (1 - \phi_2) \end{array} \Bigg )\quad \mbox{and} \quad  \textbf{f} = \left( \begin{array}{c}
0  \\ F  \end{array} \right ).\label{MCKmatrices}
\end{equation}
Using Eqn.~\eqref{MCKmatrices} we get the following expressions for the unperturbed strains $V_1$ and $V_2$,
\begin{subequations}
\label{vexact}
\begin{align}
V_1 &= F \frac{b(\phi_2)-d(\phi_2) }{a(\phi_1)b(\phi_2)+ \omega^2d(\phi_2)}\label{vexact1}\\
V_2 &= F \frac{a(\phi_1)+\omega^2 }{a(\phi_1)b(\phi_2) + \omega^2d(\phi_2)}\label{vexact2},
\end{align}
\end{subequations}
where, 
\begin{align}
a(\phi_1) = \omega_n^2(1-\phi_1)-2\omega^2 + i2\zeta\omega\omega_n, \quad b(\phi_2)= \omega_n^2(1-\phi_2)-\omega^2 + i2\zeta\omega\omega_n\notag\\ 
\mbox{and}\qquad d(\phi_2)=\omega_n^2(1-\phi_2)(1-\mu)-\omega^2+i2\zeta\omega\omega_n(1-\mu).\label{abd}
\end{align}
Next, we divide Eqs.\eqref{avgalpha02_2DOF} and~\eqref{avgalpha01_2DOF} to get
\begin{equation}\label{dphi1phi2}
\frac{d\phi_2}{d\phi_1}= \left(\frac{\phi_2}{\phi_1}\right)^{-q/p} \left\vert\frac{V_2}{V_1}\right\vert^{2/p} =\frac{\phi_2^{-q/p} | a(\phi_1)+\omega^2|^{2/p}}{\phi_1^{-q/p}| b(\phi_2)-d(\phi_2) |^{2/p}}.
\end{equation}
Since Eq.~\eqref{dphi1phi2} is easily separable, we can obtain a first integral of the averaged system. Specifically, for $q=-1$ and $p=1$, we get
\begin{multline}
\frac{\mathrm{ln}\left( \phi_1\right) \,\left( 4\omega_n^2\omega^2\zeta^2+{\omega}^{4}-2\,{\omega_n}^{2}\,{\omega}^{2}+{\omega_n}^{4}\right) }{{\mu}^{2}\,{\omega_n}^{2}}
-\mathrm{ln}\left( {\phi}_{2}\right) \,\left( 4\,{\omega}^{2}\,{\zeta}^{2}+{\omega_n}^{2}\right) +\frac{2\,{\phi}_{1}\,{\omega}^{2}}{{\mu}^{2}}\\
-\frac{\left( \left( {\phi}_{2}^{2}-4\,{\phi}_{2}\right) \,{\mu}^{2}-{\phi}_{1}^{2}+4\,{\phi}_{1}\right) \,{\omega_n}^{2}}{2\,{\mu}^{2}}=H,
\label{1stintegral}
\end{multline}
where $H$ is a constant of the motion. We then use level sets of Eq.~\eqref{1stintegral} to draw phase portraits \cite{Jordan1987} for the averaged $\phi_1$-$\phi_2$ system as shown in Fig.~\ref{Tphi1phi2}. We limit the axes in these phase-portraits to the range $[0, 1]$ as the system fails beyond this range. The shape of the trajectories are determined by the system parameters and therefore by the manner in which damage in either spring is coupled to one another. Trajectories in the phase portrait that intersect the boundary where $\phi_1=1$ fail in mode 1 and vice-versa. Further, we see that the trajectory that passes through $\phi_1=\phi_2=1$, separates the phase space into two regions - each corresponding to a different failure mode. All initial damage values on one of these two regions fail in the same failure mode. We refer to this special integral curve as the `separatrix' in the ensuing discussion.

The shape of the separatrix is determined by the system parameters. Thus, for a given set of system parameter values and an initial damage distribution, we can use shape of the separatrix to determine the probability of each failure mode. For example, we see in Fig.~\ref{Tphi1phi2} that the for different choices of parameter values, we can either have equal chances of failure in either mode (left-most plot) or most of the initial states failing in the second mode (center plot) or in the first mode (right-most plot). In the left-most plot the low forcing frequency ($\omega = 0.01$) and equal mass ratio ($\mu=1$) ensures near-equal initial strains and thus the damage evolution rate is almost the same initially. Thus the failure mode is determined \emph{largely} by the initial damage state (ie if $\phi_1\vert_{t=0} > \phi_2\vert_{t=0}$ or vice-versa) and thus the probability of failure in either mode are equal for identical distributions of the initial damage state. 

We also see the effect of coupling in the phase portraits. In the middle plot of Fig.~\ref{Tphi1phi2}, we see that for most trajectories, $\phi_1$ increases rapidly first. This implies that the displacement of the mass $m_1$ increases, since the stiffness of the first spring reduces. This in turn results in increased compression of the second spring. Since, damage in our model increases both in tension and compression, this results in a subsequent increase in $\phi_2$ and thus we see a coupling effect in the phase-portrait, wherein an increase in $\phi_1$ is followed by an increase in $\phi_2$. The final failure mode, however, can be completely determined by the position of the initial damage state in the phase portrait with respect to the separatrix.
\begin{figure}[t]
\begin{center}
\includegraphics[width=\textwidth]{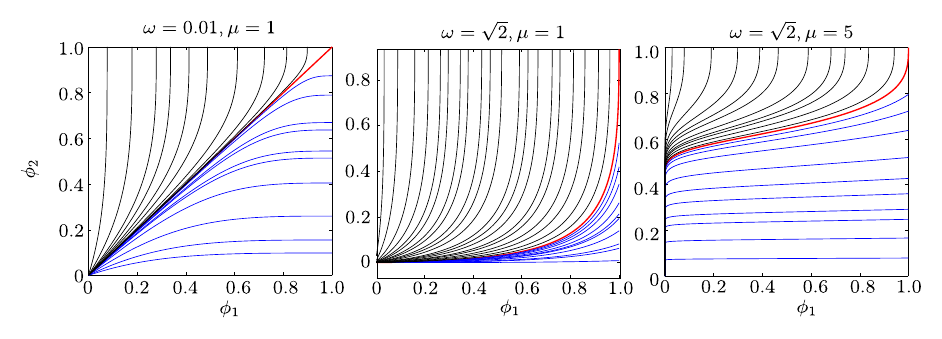}
\caption{Phase portraits of damage evolution in a 2 DOF system for different values of $\omega$ and $\mu$. Common parameter values: $\omega_n = 1.0$, $\zeta=0.003$, $p=1$, $q=-1$.}
\label{Tphi1phi2}  
\end{center}
\end{figure}

In Fig.~\ref{separatrix_position}, we show the effect of the two principal parameters, $\omega$ and $\mu$, on the shape of the separatrix (other parameters held constant). We see in Fig.~\ref{separatrix_position}(left) that for very low forcing frequencies, failure in either mode is equally likely. As the forcing frequency increases, the probability of failure in the first mode increases and then reduces. For higher frequencies the chances of failing in the second mode is much higher than that in the first mode. Fig.~\ref{separatrix_position}(right) and Fig.~\ref{Tphi1phi2}(right most) shows the effect of $\mu$ on the separatrix trajectory. If the mass ratio $\mu$ is increased, then damage initially increases only in the second spring --- since the strain in the first spring is lower due to its high inertia (that leads to lower displacement of first mass). Thereafter, due to the effect of coupling, damage in the first spring starts to increase rapidly. In the next section we use these insights gained from the damage evolution phase-portraits to analyze the failure statistics of this system.

\begin{figure}[t]
\begin{center}
\includegraphics[width=\textwidth]{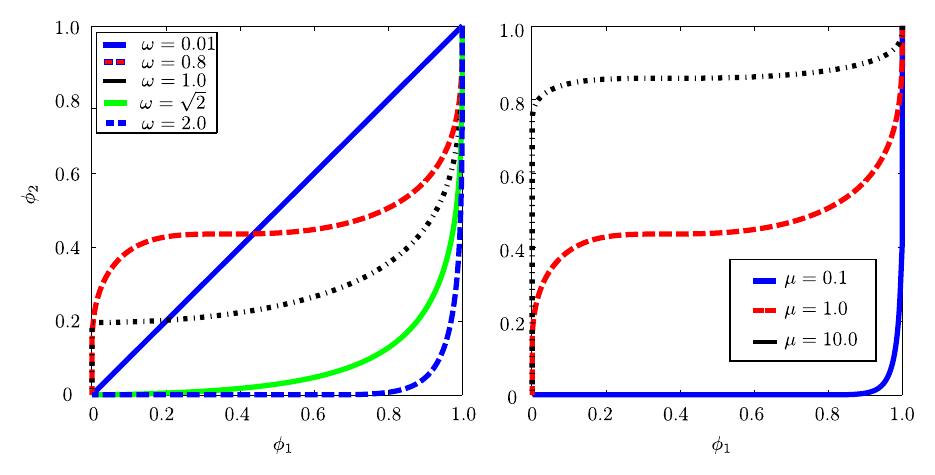}
\caption{Effect of system parameters on the shape of the separatrix. In the left plot $\omega$ is varied while $\mu = 1.0$. In the right plot $\omega = 0.8$, while $\mu$ is varied. Other parameters are the same as in Fig. \ref{Tphi1phi2}}
\label{separatrix_position}  
\end{center}
\end{figure}


\section{Failure statistics of a 2 DOF system}
\label{Failurestats}

We next use the averaged damage evolution equations to study failure statistics of the 2 DOF system. In this case we study the statistics of both failure time and location (failure mode). For this we assume that $\phi_{10} (= \phi_1(t=0))$ and $\phi_{20} (= \phi_2(t=0))$ are independent  random variables that are both uniformly distributed over a small positive interval, say $[0,\;\phi_{\max}]$. For simplicity, we perform this analysis for only two sets of damage-law exponents ($p$ and $q$), viz. when $p=1/2, q=-1$ (i.e. at the brittle limit) and $p=1, q = -1$ (i.e. away from the brittle limit). This choice of damage-exponent parameters makes it easy to derive analytical approximations to the failure-statistics which help in interpreting the results. In either case we derive closed form approximations to the failure time and location distributions and compare them to those from Monte-Carlo simulations of the averaged 2 DOF system for an ensemble of initial damage values.

\subsection{At the brittle limit}
\label{2Dbrittlelimit}

In this section we analyze the failure statistics when $p=1/2 \ \mbox{and}\  q=-1$. As discussed in section~\ref{sec:Averaging}, when the parameters $p$ and $q$ are close to the brittle limit, the \emph{unperturbed} steady state solution stays  close to the \emph{undamaged} solution for most of the life of the system. Thus, in this case, we can perform averaging by substituting the undamaged solution of the macroscopic strains $V_{1u}$ and $V_{2u}$ for $V_1$ and $V_2$.

More importantly, since the coupling between the averaged evolution laws of $\phi_1$ and $\phi_2$ is via the macroscopic strains, $V_i$, this implies that in the brittle limit $\dot{\phi_1}$ and $\dot{\phi_2}$ are not coupled to each other (as $V_{1u}$ and $V_{2u}$ are not functions of $\phi_1$ or $\phi_2$). As a result, in the brittle limit we can integrate the two damage evolution laws (Eqs.~\eqref{rescavgalpha0_2DOF}) independently to find the failure time of each spring. Specifically, for $p=1/2$ and $q=-1$,  we can determine a closed form expression for the failure time of each spring, $T_i(i=1,2)$ as follows:

\begin{equation}
\label{FailureTimeEachSpring}
T_i = \frac{1}{|V_{iu}|^4}\left(\frac{1}{\phi_{i0}}-1\right)\approx \frac{1}{|V_{iu}|^4}\frac{1}{\phi_{i0}}\qquad i=1,2,
\end{equation}
where the value of $V_{iu}$ is dependent on system parameters $\mu, \omega_n, \zeta$ and $\omega$.
Using Eq.~\eqref{FailureTimeEachSpring} and the distribution of $\phi_{i0}$, we can derive an expression for the failure time distribution for the $i$-th spring as follows:
\begin{equation}
   F_{T_i}(t) = \left\{
     \begin{array}{lr}
       0 &\; \left(t < \frac{1}{\phi_{\max}|V_{iu}|^4}\right)\\
       1 - \dfrac{1}{t\phi_{\max}|V_{iu}|^4} & \;\left(t \ge \frac{1}{\phi_{\max}|V_{iu}|^4}\right)
     \end{array}
   \right.. 
   \label{CDFTi}
\end{equation}
Now the failure time of the system, $T_f$ is given by the failure time of the spring that fails first. In other words, $T_f = \min\left(T_1, T_2\right)$. Thus the distribution of $T_f$ is given by:
\begin{align}
\label{Tfdistribution}
F_{T_f}(t)=P\left(T_f < t\right) & = P\left( \min\left(T_1,T_2\right) <t \right)\notag\\
& = 1-P\left( \min\left(T_1,T_2\right) \ge t \right)\notag\\
& = 1-P\left( T_1\ge t ,T_2\ge t \right)\notag\\
& = 1-P\left( T_1\ge t\right)P\left( T_2\ge t \right)\quad(\mbox{since $T_1$ is independent of $T_2$}) \notag\\
& = 1-\left(1-P\left( T_1< t\right)\right)\left(1-P\left( T_2<t \right)\right)\notag\\
& = 1 - \left(1-F_{T_1}(t)\right)\left(1-F_{T_2}(t)\right)\notag\\
& = \left\{
     \begin{array}{lr}
       0 &\; t < \frac{1}{\phi_{\max}V_{\max}^4}\\
       1 - \frac{1}{t\phi_{\max}V_{\max}^4} & \;\frac{1}{\phi_{\max}V_{\max}^4}<t \le \frac{1}{\phi_{\max}V_{\min}^4}\\
1-\frac{1}{t^2\phi_{\max}^2V_{\min}^4V_{\max}^4}& \;t > \frac{1}{\phi_{\max}V_{\min}^4}
     \end{array}
   \right.,
\end{align}
where $V_{\max} = \max(|V_{1u}|, |V_{2u}|)$ and $V_{\min} = \min(|V_{1u}|, |V_{2u}|)$. In Fig.~\ref{Tfail_Prob_sketch} (left), we see a comparison of the failure time distributions of $T_1$, $T_2$ and $T_f$. We also observe from Fig.~\ref{Tfail_Prob_sketch} and Eq.~\eqref{Tfdistribution}, that $\log(1-F_{T_f}) = \log(1-F_{T_1}) + \log(1-F_{T_2})$.

\begin{figure}[t]
\begin{center}
\includegraphics[width=\textwidth]{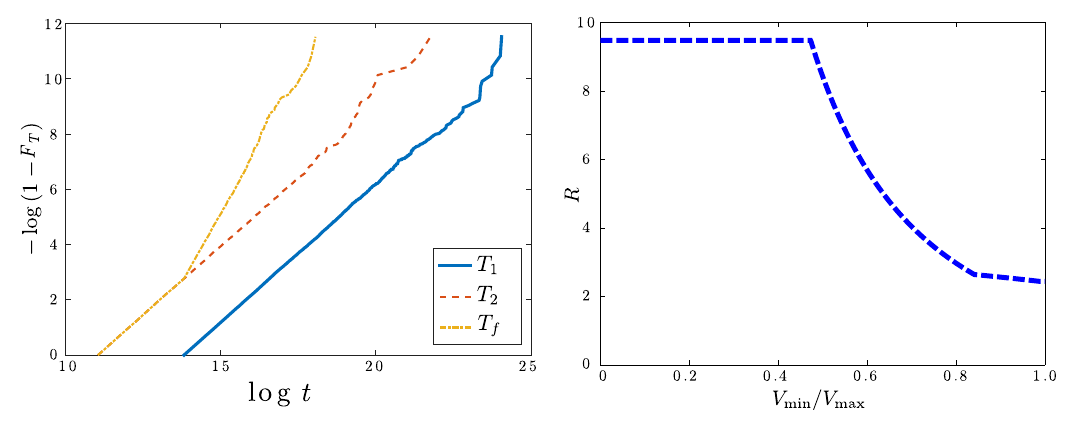}
\caption{(left) Failure Time CDF in log scale of the 2 DOF system at the brittle limit ($p=1/2 \ \mbox{and}\  q=-1$) and (right) spread of failure time, $R$, vs the undamaged strain ratio $V_r$.}
\label{Tfail_Prob_sketch}  
\end{center}
\end{figure}

We next calculate the relative failure time variability (to estimate scatter in failure times) as defined in section~\ref{FTStats}. For this we first use Eq.~\ref{Tfdistribution} to obtain an expression for the $r$-th percentile failure time $T_{[r]}$ as follows:
\begin{align}
\label{rthpercentile}
T_{[r]} &:= P\left(T_f < T_{[r]}\right)=r\\
\Rightarrow T_{[r]} &= \left\{
     \begin{array}{lr}
\frac{1}{ (1-r)\phi_{\max}V_{\max}^4} &\;\frac{V_{\min}}{V_{\max}} < \sqrt[4]{1-r}\\
\frac{1}{ \sqrt{1-r}\phi_{\max}V_{\max}^2V_{\min}^2} &\;\frac{V_{\min}}{V_{\max}} \ge \sqrt[4]{1-r}
    \end{array}
   \right.,
\end{align}
Using Eq.~\ref{rthpercentile} and Eq.~\eqref{R}, we calculate the failure time variability $R$.

\begin{equation}
R = \left\{
     \begin{array}{lr}
\frac{180}{19} &\;\frac{V_{\min}}{V_{\max}} < \sqrt[4]{0.05}\\
\frac{0.5}{ \sqrt{0.05}}\frac{V_{\max}^2}{V_{\min}^2}- \frac{10}{19}&\;\sqrt[4]{0.05} < \frac{V_{\min}}{V_{\max}} \le \sqrt[4]{0.5}\\
\sqrt{10}-\frac{\sqrt{0.5}}{0.95}\frac{V_{\min}^2}{V_{\max}^2} &\;\sqrt[4]{0.5} < \frac{V_{\min}}{V_{\max}} \le \sqrt[4]{0.95}\\
\sqrt{10}-\sqrt{\frac{10}{ 19}} &\;\sqrt[4]{0.95} < \frac{V_{\min}}{V_{\max}} \\
    \end{array}
   \right..
\end{equation}

We thus see that $R$ depends only on the undamaged strain ratio $V_r = V_{\min}/V_{\max}$. Fig.~\ref{Brittle_RS}, shows a plot between $R$ and $V_r$. We see from Fig.~\ref{Brittle_RS} that the relative variability of the failure time reduces as the undamaged strains in the two springs approaches each other, (ie as $V_r\rightarrow1$). This is because as $V_r\rightarrow1$, the failure time distribution in either springs also approach each other and hence the variability in the failure time of the system thus reduces to the failure time variability of a single spring. However, in this limit the failure location variability becomes maximized as now each failure mode is equally likely. We thus next calculate the failure location probability and determine how the variability in failure location varies with $V_r$.

For this we use our understanding of the phase portraits to determine the probability of failure of each mode. We have established in section \ref{2DOFdamage} the existence of a trajectory that divides the phase space into two half spaces, each corresponding to a different failure mode. Thus, for a given probability distribution of initial damage values, we can compute the probability of each failure mode, using the density of the initial damage on either side of this curve. Specifically, if the initial damage is distributed over a region $\mathcal{R}$ with a probability density $\mu_{\phi}$ and if the separating integral curve divides $\mathcal{R}$ into regions $\mathcal{R}_1$ and $\mathcal{R}_2$ (where $\mathcal{R}_1\cup\mathcal{R}_2=\mathcal{R}$), then the probability of failure in the first and second modes ($P_1$ and $P_2$ respectively) is given by
\begin{equation}
P_1=\int_{\mathcal{R}_1} \mu_{\phi} d\Omega\quad \mbox{and} \quad P_2=\int_{\mathcal{R}_2} \mu_{\phi} d\Omega
\end{equation}
Thus, if the initial damage is uniformly distributed over $\mathcal{R}$, then we get
\begin{equation}\label{prob}
\mu_{\phi} = \frac{1}{\int_{\mathcal{R}}  d\Omega} \Rightarrow P_1=\frac{\int_{\mathcal{R}_1} d\Omega}{\int_{\mathcal{R}}  d\Omega}\quad \mbox{and} \quad  P_2=\frac{\int_{\mathcal{R}_2} d\Omega}{\int_{\mathcal{R}}  d\Omega} = 1-P_1
\end{equation}

\begin{figure}[t]
\begin{center}
\includegraphics[width=\textwidth]{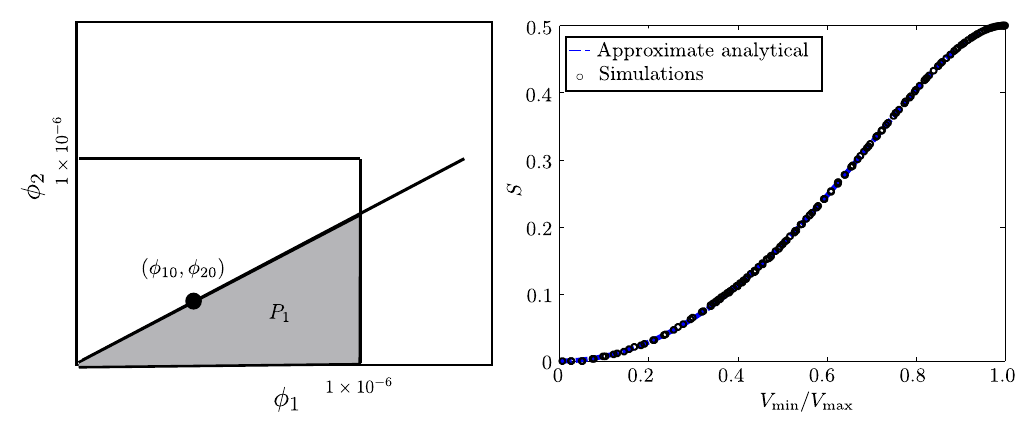}
\caption{(left) Sketch of separatrix near the initial damage state (right) S vs $V_{\min}/V_{\max}$; Parameter values for Monte-Carlo simulations: $\omega_n = 0.8$,$\zeta = 0.003$, $p=1/2$, $q=-1$, $\mu = 1$. Initial damage is uniformly distributed over $[1, 1\times10^{-6}]$ for either damage variable}
\label{Brittle_RS}  
\end{center}
\end{figure}

Thus the probability of failure of each mode can be calculated using the area of the region that the separating integral curve divides the square in Fig.~\ref{Tfail_Prob_sketch} into.
To calculate this probability, we first separate variables and integrate Eq.~\eqref{dphi1phi2} (for $p=1/2$ and $q=-1$). Specifically, if the initial damage state lies on the separatrix, then we use the fact that the separatrix passes through $\phi_1=\phi_2=1$ to obtain
\begin{equation}\label{int_dphi1phi2_brittle}
\int_{\phi_{10}}^{1}\frac{| a(\phi_1)+\omega^2|^{4}}{\phi_1^{2}}d\phi_1= \int_{\phi_{20}}^{1}\frac{| b(\phi_2)-d(\phi_2) |^{4}}{\phi_2^{2}}d\phi_2,
\end{equation}
where, $\phi_{10}$ and $\phi_{20}$ are the initial damage states. Now, the numerator in the integrand on either side is a polynomial in $\phi_i,\ i=1,2$ and  we can thus expand the numerators in Eq.~\eqref{int_dphi1phi2_brittle} to get
\begin{equation}\label{int_dphi1phi2_expanded}
\int_{\phi_{10}}^{1}\frac{A_0 + A_1\phi_1 + A_2\phi_1^2 + A_3\phi_1^3 + A_4\phi_1^4}{\phi_1^{2}}d\phi_1= \int_{\phi_{20}}^{1}\frac{B_0 + B_1\phi_2 + B_2\phi_2^2 + B_3\phi_2^3 + B_4\phi_2^4}{\phi_2^{2}}d\phi_2,
\end{equation}
where $A_i \,\mbox{and}\, B_i, \ i=0\ldots4$ are the coefficients obtained upon expanding the numerators on the left and right side of Eq.~\eqref{int_dphi1phi2_brittle} respectively. Now, since $0\le\phi_{10},\phi_{20}\le\phi_{\max}\ll1$, if we integrate both sides in Eq.~\eqref{int_dphi1phi2_brittle}, we get
\begin{equation}\label{phi20byphi10_brittle_1}
\frac{A_0}{\phi_{10}}    = \frac{B_0}{\phi_{20}} + \mathcal{O}(1).
\end{equation}
Substituting values for $A_0$ and $B_0$ in Eq.~\eqref{phi20byphi10_brittle_1}, we get
\begin{align}\label{phi20byphi10_brittle}
\frac{| a(\phi_1=0)+\omega^2|^4}{\phi_{10}}    & = \frac{|b(\phi_2=0)-d(\phi_2=0) |^4}{\phi_{20}} + \mathcal{O}(1) \notag\\
\Rightarrow \frac{\phi_{20}}{\phi_{10}}& = \left\vert\frac{b(\phi_2=0)-d(\phi_2=0) }{a(\phi_1=0)+\omega^2}\right\vert^4 + \mathcal{O}(\phi_{20})= \left\vert\frac{V_{1u}}{V_{2u}}\right\vert^4 + \mathcal{O}(\phi_{20}).
\end{align}
We next use Eqs.~\eqref{dphi1phi2} and ~\eqref{phi20byphi10_brittle} (retaining terms only till $\mathcal{O}(1)$) to calculate the slope of the separatrix at the initial state $(\phi_{10},\phi_{20})$ as:
\begin{equation}
\label{dphi1dphi2_brittle}
\left.\frac{d\phi_{2}}{d\phi_{1}}\right\vert_{\begin{subarray}{l}\phi_1 = \phi_{10}\\
    \phi_2 = \phi_{20}\end{subarray}} = \left.\left(\frac{\phi_2}{\phi_1}\right)^{2} \left\vert\frac{V_2}{V_1}\right\vert^{4}\right\vert_{\begin{subarray}{l}\phi_1 = \phi_{10}\\
    \phi_2 = \phi_{20}\end{subarray}}  \approx \left(\frac{\phi_{20}}{\phi_{10}}\right)^2\left\vert\frac{V_{2u}}{V_{1u}}\right\vert^4  \approx \frac{\phi_{20}}{\phi_{10}}\quad \mbox{(using Eq.~\eqref{phi20byphi10_brittle})}.
\end{equation}

Note, that for this case we approximated the unperturbed strains at the initial damage values in Eqs.~\eqref{dphi1phi2}, $V_{i},\ i=1,2$, with the undamaged strains $V_{iu},\ i=1,2$. Eq.~\eqref{dphi1dphi2_brittle} implies that close to the small initial damage values, the separatrix is a straight line passing through the origin. It is important to note that $\phi_1 =\phi_2 = 0$ is a fixed point in the averaged damage evolution space and hence no separatrix would actually pass through it. The result obtained in Eq.~\eqref{dphi1dphi2_brittle} is a consequence of the assumptions made during deriving this approximate expression. This approximation, however, is adequate for calculating failure location probabilities as indicated in the ensuing results.

We use Eq.~\eqref{dphi1dphi2_brittle} to calculate the area of the two regions on either side of the separatrix (as shown in Fig.~\ref{Brittle_RS}) and therefore calculate $P_1$ and $P_2$ using Eq.~\eqref{prob}, as follows:

\begin{equation}
P_1 = \left\{
     \begin{array}{lr}
\frac{1}{2}\left\vert\frac{V_{1u}}{V_{2u}}\right\vert^4 &\;\frac{|V_{1u}|}{|V_{2u}|} < 1\\
1-\frac{1}{2}\left\vert\frac{V_{2u}}{V_{1u}}\right\vert^4&\;\frac{|V_{1u}|}{|V_{2u}|} \ge 1\\
    \end{array}
   \right.,\qquad
P_2 = 1-P_1
\end{equation}
Since the failure location is a random variable following the Bernoulli distribution \cite{BernoulliDistr} (the two states being the two failure modes) with parameter $P_1$, we can quantify the uncertainty in failure location, $S$, by the standard deviation of the Bernoulli distribution, failure location:
\begin{equation}
\label{S}
S = \sqrt{{P_1}{P_2}} =  \sqrt{\frac{1}{2}\left\vert\frac{V_{\min}}{V_{\max}}\right\vert^4\left(1-\frac{1}{2}\left\vert\frac{V_{\min}}{V_{\max}}\right\vert^4  \right)}.
\end{equation}
Thus the failure location uncertainty depends solely on the undamaged strain ratio $V_r = V_{\min}/V_{\max}$ and not their individual values. In Fig.~\ref{Brittle_RS} (right), we see a comparison of the analytical approximation of $S$ as given by Eq.~\eqref{S} to that obtained through numerical simulations of the averaged damage evolution laws. We also see that the uncertainty in failure location increases monotonically with $V_r$. This is because in the brittle limit, the damage growth in the two springs becomes decoupled and thus as $V_r\rightarrow1$, either failure mode becomes  equally likely. Thus, by comparing Figs.~\ref{Tfail_Prob_sketch} (right) and ~\ref{Brittle_RS} (right), we see that as $V_r\rightarrow1$, the uncertainty in failure location increases while the uncertainty in failure time reduces.

\subsection{Away from the brittle limit}

\begin{figure}[t]
\begin{center}
\includegraphics[width=\textwidth]{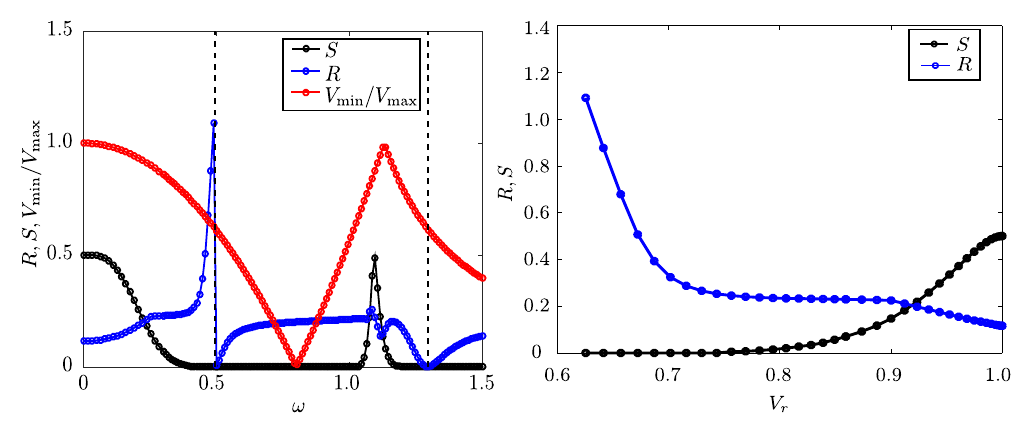}
\caption{(left) $R$, $S$, $V_{\min}/V_{\max}$ vs $\omega$; (right) $S$ vs $V_r (=V_{\min}/V_{\max})$ when $0<\omega<0.5$. Parameter values: $\omega_n = 0.8$, $\zeta = 0.003$, $p=1$, $q=-1$, $\mu = 1$. Initial damage is uniformly distributed over $[1, 1\times10^{-6}]$ for either damage variable.}
\label{RS_ductile}  
\end{center}
\end{figure}

We next study the failure statistics in the limit when damage growth in either spring is coupled to one another. However, as a result of the coupled damage evolution in this case, it is difficult to obtain a closed form approximation to the failure time distribution and we thus rely on simulations to study failure time statistics. Specifically, we choose parameter values $\omega_n = 0.8$, $\zeta = 0.003$, $p=1$, $q=-1$, $\mu = 1$ and then simulate damage evolution for an ensemble of initial damage distributed over $[1, 1\times10^{-6}]$ for $\phi_1$ and $\phi_2$. For each case we record the failure time and mode and use that to compute $R$ and $S$ as in section \ref{2Dbrittlelimit}.

In Fig.~\ref{RS_ductile}(left), we plot how the variability in failure time and failure location varies with the forcing frequency. We also plot how the undamaged strain ratio $V_r$, varies with $\omega$ in this case. We observe that there are regions in the frequency spectrum wherein $R$ monotonically increases and $S$ monotonically decreases as observed in the brittle limit. This can also be seen in Fig.~\ref{RS_ductile}(right), wherein $R$ and $S$ for $0<\omega<0.5$ is plotted versus $V_r$. However, it is evident from Fig.~\ref{RS_ductile}(left) that $R$ and $S$ are not just a function of the undamaged strain ratio $V_r$. Also, that the monotonic behavior observed in the brittle limit, although true for some ranges of the forcing frequency, does not hold true for all forcing frequencies in general.


\section{Conclusion}
In this paper, we used the perturbation method of averaging to derive averaged damage evolution rate laws from 1 and 2 DOF coupled-field damage models. While in both cases the macroscopic component of the models is linear for constant damage, the overall system structures are nonlinear due to the coupling between the microscopic damage and macroscopic displacement configuration variables. Through this approach, we were able to project the macroscopic dynamics occurring at the fast time scale onto the much slower time scale of damage evolution. Because it permits numerical solutions to march along the slower time scale of the microscopic damage evolution, this substantially reduced life-cycle simulation times, in this study by an order of magnitude. In applications, this speed up can be expected to be by many orders of magnitude, depending on the size of the damage rate constant. We then used the averaged damage evolution laws to study the effect of system parameters on damage growth and failure statistics. While this paper focused on the derivation of rate laws for 1 and 2 DOF model systems, the same approach can, in principle, be extended to higher dimensional systems.

In addition to greatly improving computational efficiency, the averaged equations make possible analyses that are otherwise very difficult, if not impossible. For the 1 DOF system we used the averaged rate laws to determine regimes of the damage growth exponent for which the system behaves like a brittle material (i.e., exhibiting nearly zero damage growth followed by sudden large damage growth towards the end of life). We also derived expressions for the failure time distributions for certain specific choices of the damage growth exponent. With these expressions, we were able to understand how system parameters affect failure time variability. We also compared these analytical results against Monte-Carlo life-cycle simulations for large ensembles of random initial damage states, which was possible because the averaged equations so significantly reduced the simulation time for each individual simulation by an order of magnitude. We demonstrated that the variability in failure times is higher for systems which behave like a brittle material. We also used the averaged equations to derive an expression for the $F$-$N$ curve (the equivalent for lumped models to the $S$-$N$ curve of distributed systems): it was demonstrated that the statistical variability in the $F$-$N$ curves is the same at all forcing levels.

We next derived the averaged damage rate laws for a 2 DOF system. The additional DOF results in a new failure mode and, thus, this model is among the simplest permitting the study of both failure times and locations. We constructed damage evolution phase portraits for different system parameter values, and demonstrated the existence of a damage separatrix  that divides the damage phase space into two subsets, each containing initial damage states the give trajectories terminating in one failure mode or the other. Thus, the shape of the separatrix for a given set of parameter values, together with the probability distribution of initial damage states, can be used to determine the probability of failure at each location. Using a similar approach to that used for the 1 DOF model, we then obtained approximate expressions for the failure time and location probability distributions. We demonstrated using these expressions that, for materials with brittle-like behavior, the relative variability in failure time and location depend only on the ratio of the undamaged macroscopic strains of the two springs of the 2 DOF system. We also demonstrated an inverse relationship between temporal and spatial variability: the variability in failure time decreases as the variability in failure location increases. Finally, we showed via simulations that, for other choices of the damage evolution exponent, this result holds true, at least for certain ranges of the forcing frequency.

\bibliographystyle{unsrtnat}
\bibliography{Main_database_2023}

\end{document}